\documentclass[12pt,epsf]{article}
\usepackage{graphicx}
\usepackage{epsfig}
\begin{document}

\def\a{\alpha}
\def\b{\beta}
\def\c{\varepsilon}
\def\d{\delta}
\def\e{\epsilon}
\def\f{\phi}
\def\g{\gamma}
\def\h{\theta}
\def\k{\kappa}
\def\l{\lambda}
\def\m{\mu}
\def\n{\nu}
\def\p{\psi}
\def\q{\partial}
\def\r{\rho}
\def\s{\sigma}
\def\t{\tau}
\def\u{\upsilon}
\def\v{\varphi}
\def\w{\omega}
\def\x{\xi}
\def\y{\eta}
\def\z{\zeta}
\def\D{\Delta}
\def\G{\Gamma}
\def\H{\Theta}
\def\L{\Lambda}
\def\F{\Phi}
\def\P{\Psi}
\def\S{\Sigma}

\def\o{\over}
\def\beq{\begin{eqnarray}}
\def\eeq{\end{eqnarray}}
\newcommand{\gsim}{ \mathop{}_{\textstyle \sim}^{\textstyle >} }
\newcommand{\lsim}{ \mathop{}_{\textstyle \sim}^{\textstyle <} }
\newcommand{\vev}[1]{ \left\langle {#1} \right\rangle }
\newcommand{\bra}[1]{ \langle {#1} | }
\newcommand{\ket}[1]{ | {#1} \rangle }
\newcommand{\EV}{ {\rm eV} }
\newcommand{\KEV}{ {\rm keV} }
\newcommand{\MEV}{ {\rm MeV} }
\newcommand{\GEV}{ {\rm GeV} }
\newcommand{\TEV}{ {\rm TeV} }
\def\diag{\mathop{\rm diag}\nolimits}
\def\Spin{\mathop{\rm Spin}}
\def\SO{\mathop{\rm SO}}
\def\O{\mathop{\rm O}}
\def\SU{\mathop{\rm SU}}
\def\U{\mathop{\rm U}}
\def\Sp{\mathop{\rm Sp}}
\def\SL{\mathop{\rm SL}}
\def\tr{\mathop{\rm tr}}

\def\IJMP{Int.~J.~Mod.~Phys. }
\def\MPL{Mod.~Phys.~Lett. }
\def\NP{Nucl.~Phys. }
\def\PL{Phys.~Lett. }
\def\PR{Phys.~Rev. }
\def\PRL{Phys.~Rev.~Lett. }
\def\PTP{Prog.~Theor.~Phys. }
\def\ZP{Z.~Phys. }


\baselineskip 0.7cm

\begin{titlepage}

\begin{flushright}
UT-08-31\\
IPMU-08-0088
\end{flushright}

\vskip 1.35cm
\begin{center}
{\large \bf
   Decaying Dark Matter Baryons\\ in a Composite Messenger Model
}
\vskip 1.2cm
Koichi Hamaguchi$^{1,2}$, Eita Nakamura$^1$, Satoshi Shirai$^1$ and T. T. Yanagida$^{1,2}$
\vskip 0.4cm

{\it $^1$  Department of Physics, University of Tokyo,\\
     Tokyo 113-0033, Japan\\
$^2$ Institute for the Physics and Mathematics of the Universe, 
University of Tokyo,\\ Chiba 277-8568, Japan}

\vskip 1.5cm

\abstract{
A baryonic bound state with a mass of ${\cal O}(100)$ TeV, which is composed of
strongly interacting messenger quarks in the low scale gauge mediation, 
can naturally be the cold dark matter.
Interestingly, we find that such a baryonic dark matter is generically metastable, 
and the decay of this dark matter can naturally explain the anomalous positron flux 
recently observed by the PAMELA collaboration.
}
\end{center}
\end{titlepage}

\setcounter{page}{2}

\section{Introduction}

The origin and nature of the dark matter (DM) are one of the most challenging problems in the particle
physics and cosmology.
If the DM is in thermal equilibrium in the early universe, the present abundance
of the DM is determined by its annihilation cross section $\sigma v_{rel}$. Here, $v_{rel}$ is 
the relative velocity of the DM at the freeze out time. The observed DM density requires 
$\sigma v_{rel} \simeq 2\times 10^{-9}\ {\rm GeV}^{-2}$. From this result one may derive 
the upper bound
on the mass of the DM as $m_{\rm DM} \lsim 100$ TeV \cite{Griest:1989wd}. This is because
the ($s$-wave) annihilation cross section never violates the unitarity limit given by
\begin{equation}
\sigma v_{rel} \lsim \frac{4\pi}{m_{\rm DM}^2v_{rel}}.
\end{equation}

The upper-bound value, $m_{\rm DM} \simeq 100$ TeV, is obtained when the annihilation cross section
saturates the unitarity limit and hence the DM is subject to a new strong interaction
at the energy scale of 100 TeV. Interestingly, the scale 100 TeV coincides with the supersymmetry (SUSY) breaking scale 
in the low scale gauge mediation.
Therefore, it is intriguing that the DM is one of composite bound states
in hidden quarks $Q$'s in the dynamical SUSY-breaking hidden sector \cite{Dimopoulos:1996gy, INMY}.

In a recent article \cite{Hamaguchi:2007rb}, we proposed DM baryons composed of 
the messenger quarks $P$'s, 
instead of the hidden quarks $Q$'s responsible for the SUSY breaking. Here, the messenger
quarks $P^i_\alpha$ and ${\bar P}_i^\alpha$ are assumed to transform
as ${\bf 5}$ and ${\bf 5^*}$ under the GUT gauge group, and 
as anti-fundamental and fundamental representations of a new 
strongly interacting gauge group ${\rm SU}(N)$,
respectively. 
Here, the subscript $i$ ($\alpha$) represents the GUT (${\rm SU}(N)$) index.
The DM baryons are expressed as 
$B^{i_1i_2\cdots i_N} = \epsilon^{\a_1\a_2\cdots \a_N} P_{\a_1}^{i_1}P_{\a_2}^{i_2}\cdots P_{\a_N}^{i_N}$ 
and ${\bar B}_{i_1i_2\cdots i_N} = \epsilon_{\a_1\a_2\cdots \a_N} {\bar P}^{\a_1}_{i_1}{\bar P}^{\a_2}_{i_2}\cdots {\bar P}^{\a_N}_{i_N}$.
The requirement that these baryons are neutral under the GUT gauge group dictates $N=5$.

If the new gauge group is indeed SU(5), the baryon-number violating operators in the superpotential
are allowed, $W= (1/M_P^2)P^5 + (1/M_P^2){\bar P}^5$, and the composite baryons become metastable.
Here, $M_P=2.4\times 10^{18}$ GeV is the reduced Planck scale.
We find that the lifetime of the baryon DM is ${\cal O}(10^{24-26})$ sec. 
It is much longer than the age of the universe and assures the stability of the DM, while
it is short enough to leave observable signatures in cosmic rays.
Especially, the anomalous positron flux 
recently observed by the PAMELA collaboration~\cite{Adriani:2008zr} is naturally explained by the 
decay of the composite DM with the above mass and lifetime.

\section{Strongly interacting messenger model and composite baryons}

We consider a SUSY extension of the standard model (SM) in which a SUSY-breaking effect of a hidden sector
is communicated to the visible minimal supersymmetric standard model (MSSM) sector by the gauge mediation
mechanism~\cite{Giudice:1998bp}. 
We represent the SUSY-breaking effect of the hidden sector as nonzero vacuum expectation values
of scalar and $F$ components of a singlet field $\Phi_S$ belonging to the hidden sector: $\langle\Phi_S\rangle=\langle\varphi_S\rangle
+\theta^2\langle F_S\rangle$. We introduce messengers $P^i_\alpha=(L_\alpha,D_\alpha)$ and 
$\bar{P}_i^\alpha=(\bar{L}^\alpha,\bar{D}^\alpha)$ which are in the
fundamental ($\bf5$) and anti-fundamental ($\bf5^*$) representations of the
SM gauge group ${\rm SU}(5)_{\rm GUT}\supset{\rm SU}(3)\times{\rm SU}(2)\times{\rm U}(1)$.
Here, $i=1,\cdots,5$ is the index for ${\rm SU}(5)_{\rm GUT}$.
We assume that the messengers are also in anti-fundamental ($\bf5^*$) and fundamental ($\bf5$) representations of
 an additional gauge group ${\rm SU}(5)$ as explained in the Introduction (see also Ref.~\cite{Hamaguchi:2007rb}). 
The above introduced subscript $\alpha=1,\cdots,5$ represents
the index for this gauge group. A crucial assumption here is that this additional gauge group
is strongly coupled at the messenger mass scale.

The messengers interact with the hidden sector via a Yukawa interaction\footnote{In general
the Yukawa couplings of $D$ and $L$ can be different, i.e.,
$W = 
y_L \Phi_S L_\alpha {\bar L}^\alpha 
+
y_D \Phi_S D_\alpha {\bar D}^\alpha$.}
\begin{equation}
 W=y\Phi_SP^i_\alpha\bar{P}_i^\alpha,
\end{equation}
while they interact with the MSSM sector via the SM gauge interaction.
The nonzero scalar and $F$ components of the $\Phi_S$ yield a mass term for the messengers.
This setup is a standard form of the gauge mediation~\cite{Giudice:1998bp}.

We now turn to the strongly interacting ${\rm SU}(5)$ sector. This sector is a SUSY 
QCD with the number of fundamentals equal to the rank of the gauge group ${\rm SU}(5)$ 
(i.e., $N_f = N = 5$).
The low energy effective theory of this gauge theory is described by a confined theory with
a deformed moduli space~\cite{Seiberg}. This theory can be described by ${\rm SU}(5)$
gauge invariant operators made up
of messenger quarks. They are mesons
\begin{equation}
 M^i_j=P^i_\alpha\bar{P}^\alpha_j 
\end{equation}
and baryons
\begin{equation}
 B=P^5=\frac{1}{5!}\epsilon_{ijklm}\epsilon^{\alpha\beta\gamma\delta\epsilon}
P^i_\alpha P_\beta^j P_\gamma^k P_\delta^l P_\epsilon^m,\quad
\bar{B}=\bar{P}^5=\frac{1}{5!}
\epsilon^{ijklm}\epsilon_{\alpha\beta\gamma\delta\epsilon}
\bar{P}_i^\alpha \bar{P}^\beta_j \bar{P}^\gamma_k \bar{P}^\delta_l \bar{P}^\epsilon_m.
\end{equation}
The effective superpotential is given by (with a Lagrange's multiplier field $X$)~\cite{Seiberg},
\begin{equation}
W=X({\rm det}\,M-B\bar{B}-\Lambda^{10})+m_i^jM^i_j,
\end{equation}
where, for simplicity, we consider a supersymmetric mass term for messengers.
This term and non-supersymmetric mass terms for messengers are generated by the 
Yukawa term with the vacuum expectation value of $\Phi_S$.
This superpotential yields a SUSY invariant vacuum at
\begin{eqnarray} 
\langle M^i_j\rangle &=& [m^{-1}]_j^i(\Lambda^{10}{\rm det}\,m)^{\frac{1}{5}}, \\
\langle B\rangle &=& \langle\bar{B}\rangle\ =\ 0, \\
\langle X\rangle &=& -\frac{(\Lambda^{10}{\rm det}\,m)^{\frac{1}{5}}}{\Lambda^{10}}.
\end{eqnarray}

We can also estimate the masses of the baryons and mesons of the confined theory.
For this purpose, however, we must know the form of the K\"ahler potential, in particular, the normalization of the kinetic terms.
We estimate this by using the Naive Dimensional Analysis (NDA) method \cite{Luty}, which assumes that the confined theory
is strongly coupled. The NDA method gives the above superpotential in terms of canonically normalized fields $\hat{B},\
\hat{\bar{B}}$, $\hat{M}$ and $\hat{X}$ as
\begin{eqnarray}
W &=& \frac{1}{g^2}\left[\frac{g\hat{X}}{\Lambda^8}\left({\rm det}(g\Lambda\hat{M})-g^2\Lambda^8\hat{B}\hat{\bar{B}}-\Lambda^{10}\right)
+m_i^j(g\Lambda\hat{M}^i_j)\right] \\
&=& \hat{X}\left(\frac{g^4}{\Lambda^3}{\rm det}\,\hat{M}-g\hat{B}\hat{\bar{B}}-\frac{1}{g}\Lambda^2\right)+\frac{\Lambda}{g}m_i^j\hat{M}_j^i,
\end{eqnarray}
where $g$ is a constant of order $4\pi$.
The corresponding vacuum expectation values are
\begin{eqnarray}
\langle \hat{M}^i_j\rangle &=& [m^{-1}]_j^i\frac{(\Lambda^{5}{\rm det}\,m)^{\frac{1}{5}}}{g}, \\
\langle \hat{B}\rangle &=& \langle\hat{\bar{B}}\rangle\ =\ 0,\label{eq:B} \\
\langle \hat{X}\rangle &=& -\frac{(\Lambda^{5}{\rm det}\,m)^{\frac{1}{5}}}{g\Lambda}.
\end{eqnarray}
We can see from the above expressions that both the baryon and meson masses are of order $m$.

The baryon and anti-baryon in this theory are neutral under the SM gauge group.
Note also that the baryon number is automatically conserved by the superpotential and 
the vacuum expectation values. Assuming the baryon number conservation in the 
K\"ahler potential, the (anti-)baryons become stable and can be candidates
for the DM.
Now if the baryons and mesons are really strongly coupled, the annihilation cross sections of the
(anti-)baryons nearly saturate the unitarity limit, and we must have $m\sim100~{\rm TeV}$ to explain the observed
DM density of the universe as discussed in the Introduction. It is very interesting that this mass scale
coincides with the messenger mass scale appearing in the gauge mediation, and
we have a good reason to identify these baryons of the messenger sector as
a candidate for the DM.

\section{Possible baryon-number violating operators and decays of the DM baryons}

Although we have baryon number conservation to leading order in the messenger sector, 
it can be violated by higher dimensional operators.
These baryon-number violating effects can be represented by Planck-suppressed higher dimensional operators in
the superpotential. The leading terms are
\begin{equation}
\Delta W=\frac{1}{M_P^2}P^5+\frac{1}{M_P^2}\bar{P}^5
\;\to\;
\frac{1}{M_P^2}\frac{\Lambda^4\hat{B}}{g}+\frac{1}{M_P^2}\frac{\Lambda^4\hat{\bar{B}}}{g},
\end{equation}
where we omit unknown coefficients of order unity.
In the confined description, these terms are linear in $\hat{B}$ and $\hat{\bar{B}}$, and therefore
produce a slight shift in the vacuum expectation values of them.
We have, instead of Eq.~(\ref{eq:B}),
\begin{equation}
\langle \hat{B}\rangle\simeq\langle\hat{\bar{B}}\rangle\simeq-\frac{\Lambda^5}{gM_P^2(\Lambda^5{\rm det}\,m)^{\frac{1}{5}}}.
\end{equation}

There are interactions between baryons and mesons coming from solving the constraining equation.
Upon integrating $\hat{X}$ and expanding around the vacuum expectation values,
\begin{equation}
W\ni \frac{gm}{\Lambda}\hat{B}\hat{\bar{B}}\hat{M},\quad \frac{g^2m}{\Lambda^{2}}\hat{B}\hat{\bar{B}}\hat{M}^2\quad
{\rm etc.}
\end{equation}
There may also be contribution coming from the K\"ahler potential, e.g.,
\begin{equation}
K\ni \frac{g^2}{\Lambda^{2}}\hat{B}^\dagger\hat{B}\hat{M}^\dagger\hat{M},\quad
\frac{g^2}{\Lambda^{2}}\hat{\bar{B}}^\dagger\hat{\bar{B}}\hat{M}^\dagger\hat{M}.
\end{equation}
These terms and the above baryon-number violating effects induce the baryons to decay into
mesons, which in turn decay into MSSM particles by the gauge interaction or
higher dimensional terms in the superpotential (see Ref.~\cite{Hamaguchi:2007rb}).
Assuming that the messenger mass $m$ is of the same order as $\Lambda$, the lifetime of
the baryons are estimated as
\begin{equation}
\tau\sim 16\pi \frac{M_P^4}{g^2\Lambda^5}\sim\frac{M_P^4}{\pi\Lambda^5},
\end{equation}
where the factor $16\pi$ comes from the phase space integral.
With $m\sim\Lambda\sim (30$--$100)\ {\rm TeV}$, this yields numerically $\tau\sim10^{24-26}\ {\rm sec}$.
Note that the theory with $N_f=N=4$ or $6$, instead of $5$,
would lead to a lifetime of $10^{-(1-3)}$ sec or 
$10^{51-54}$ sec, respectively, which shows that the prediction 
$\tau\sim10^{24-26}\ {\rm sec}$ for $N=5$ is quite nontrivial.
As we will see, this lifetime together with the DM mass can naturally explain the anomalous positron
excess observed by the PAMELA collaboration.

\section{Positrons from the DM decays}
\begin{figure}[h]
\begin{center}
\scalebox{1}{
\begin{picture}(220,241.249)(0,0)
\put(5,21.8956){\line(1,0){20}}
\put(10,24.8956){$\scriptstyle h^0$}
\put(5,82.8841){\line(1,0){20}}
\put(5,81.9871){\line(1,0){20}}
 \put(5,85.8841){$\scriptstyle H^0\, A^0$}
\put(5,84.0538){\line(1,0){20}}
 \put(5,74.0538){$\scriptstyle H^\pm$}
\put(37,41.8978){\line(1,0){20}}
\put(42.,33.8978){$\scriptstyle {\tilde{\chi}^0_1}$}
\put(37,60.0119){\line(1,0){20}}
\put(42.,52.0119){$\scriptstyle {\tilde{\chi}^0_2}$}
\put(37,63.1396){\line(1,0){20}}
\put(42.,67.1396){$\scriptstyle {\tilde{\chi}^0_3}$}
\put(37,94.8663){\line(1,0){20}}
\put(42.,98.8663){$\scriptstyle {\tilde{\chi}^0_4}$}
\put(69,58.5031){\line(1,0){20}}
\put(74.,62.5031){$\scriptstyle {\tilde{\chi}^{\pm}}_1$}
\put(69,93.8715){\line(1,0){20}}
\put(74.,97.8715){$\scriptstyle {\tilde{\chi}^{\pm}_2}$}
\put(101,241.249){\line(1,0){20}}
\put(106,245.249){$\scriptstyle {\tilde g}$}
\put(133,53.802){\line(1,0){20}}
\put(133,51.5244){\line(1,0){20}}
\put(138,43.5244){$\scriptstyle \tilde \nu_e$}
\put(138,57.802){$\scriptstyle \tilde e_L$}
\put(133,26.0498){\line(1,0){20}}
\put(138,30.0498){$\scriptstyle \tilde e_R $}
\put(165,173.832){\line(1,0){20}}
\put(170,165.832){$\scriptstyle \tilde t_1$}
\put(165,192.465){\line(1,0){20}}
\put(165,185.755){\line(1,0){20}}
\put(165,186.897){\line(1,0){20}}
\put(165,194.897){$\scriptstyle \tilde b_2\,\tilde t_2\,\tilde b_1$}
\put(165,25.9173){\line(1,0){20}}
\put(170,29.9173){$\scriptstyle \tilde \tau_1$}
\put(165,53.8107){\line(1,0){20}}
\put(170,57.8107){$\scriptstyle \tilde \tau_2$}
\put(165,51.5162){\line(1,0){20}}
\put(170,43.5162){$\scriptstyle \tilde \nu_{\tau}$}
\put(240,120.625){\small ~~Mass }
\put(240,110.625){$\scriptstyle {\rm ~~[GeV]}$ }
\put(220,-10){\vector(0,1){262.249}}
\put(220,0){\line(1,0){10}}
\put(232,-2){$\scriptstyle 0$}
\put(220,40){\line(1,0){10}}
\put(232,38){$\scriptstyle 200$}
\put(220,80){\line(1,0){10}}
\put(232,78){$\scriptstyle 400$}
\put(220,120){\line(1,0){10}}
\put(232,118){$\scriptstyle 600$}
\put(220,160){\line(1,0){10}}
\put(232,158){$\scriptstyle 800$}
\put(220,200){\line(1,0){10}}
\put(232,198){$\scriptstyle 1000$}
\put(220,240){\line(1,0){10}}
\put(232,238){$\scriptstyle 1200$}
\put(220,0){\line(1,0){5}}
\put(220,20){\line(1,0){5}}
\put(220,40){\line(1,0){5}}
\put(220,60){\line(1,0){5}}
\put(220,80){\line(1,0){5}}
\put(220,100){\line(1,0){5}}
\put(220,120){\line(1,0){5}}
\put(220,140){\line(1,0){5}}
\put(220,160){\line(1,0){5}}
\put(220,180){\line(1,0){5}}
\put(220,200){\line(1,0){5}}
\put(220,220){\line(1,0){5}}
\end{picture} }
\caption[]{MSSM mass spectrum. The gravitino mass is 0.3 eV. }
\label{fig:MSSM}
\end{center}
\end{figure}
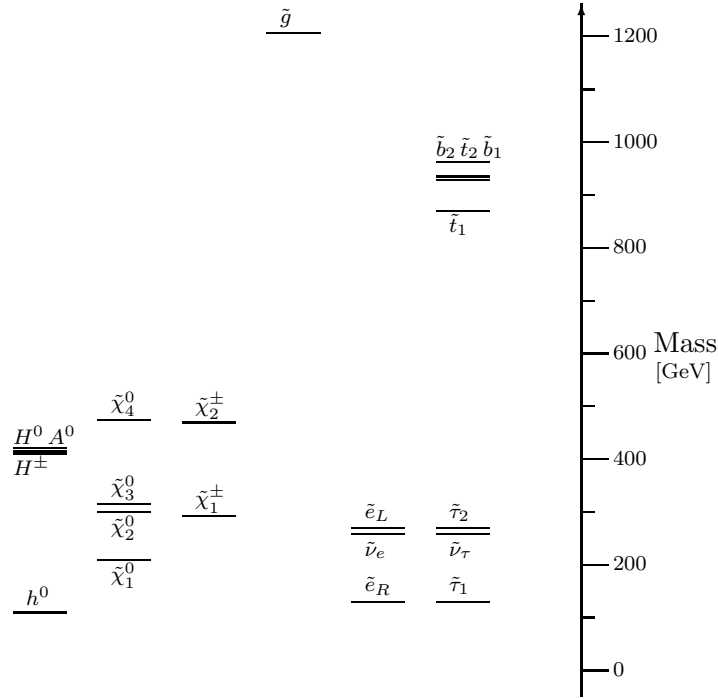

Before discussing the signals of the DM decays, we give a typical SUSY particle spectrum for concreteness. 
In our model, it is not easy to calculate the low-energy mass spectrum,
since the messenger sector is strongly interacting.
We expect that, from the viewpoint of effective theory,
the ${\rm SU}(5)_{\rm GUT}$ adjoint messenger mesons (such as $L\bar{L},D\bar{D},L\bar{D},D\bar{L}$) 
generate the MSSM sparticle masses.
Thus, for simplicity, we adopt a minimal gauge mediated SUSY breaking (mGMSB) model with $N_5=5$.
From now on, we take mGMSB parameters as
\begin{equation}
\Lambda_{\rm SUSY} = 30~{\rm TeV},~M = 40~{\rm TeV},~N_5 = 5,~\tan\beta = 5,~{\rm sgn}(\mu)=+1~{\rm and}~C_{\rm grav}=1.
\end{equation}
The mass spectrum is shown in Fig.~\ref{fig:MSSM}, which is calculated by ISAJET 7.72~\cite{ISAJET}.
 It also predicts an ultralight gravitino 
with mass 0.3 eV.

Now let us discuss the signals of the DM decays.
For simplicity, we assume that the DM baryon decays to only two lightest $L\bar{L}$ mesons.\footnote{
Because of the renormalization effect from $\SU(5)_{\rm GUT}$, the leptonic messenger tends to be light.
Therefore, it would be reasonable to assume that the baryon DM dominantly decays to $L\bar{L}$ mesons because of kinematical reason.
Here, for simplicity, we assume that the baryon DM decays into a pair of SU(2) singlet $L\bar{L}$'s,
neglecting the effects of the $D\bar{D}$ component in the singlet as well as the decay into the SU(2) triplet $L\bar{L}$'s.
}
This meson decays to a pair of gauginos or gauge bosons through triangle anomaly-like diagrams.
A precise calculation of mesons' branching fractions is difficult, because of
the strong dynamics as well as the SUSY breaking effects. 
Here, for simplicity, we assume that the mesons decay into a pair of gauge bosons or 
a pair of gauginos with the same branching fraction, as in the SUSY conserving limit.

\begin{figure}[h]
\begin{center}
\epsfig{file=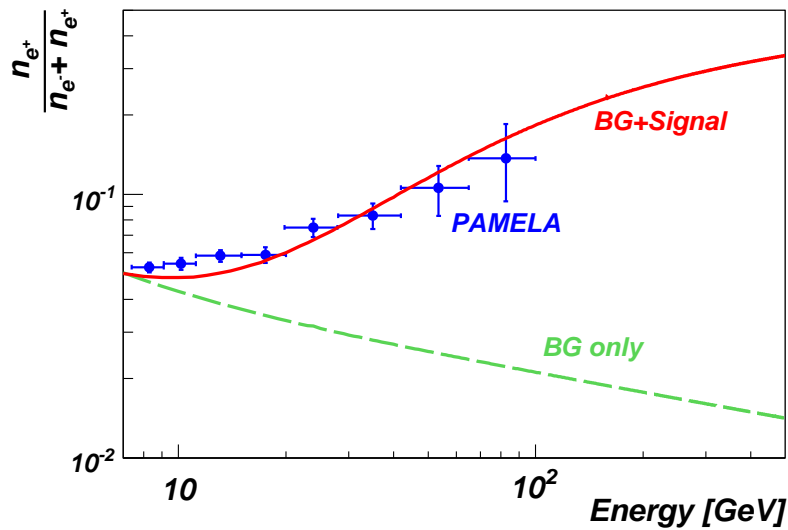,clip,width=11cm}
\caption[]{Positron fraction from decay of the baryon with $m_B=100$ TeV and $m_{M}=40$ TeV.
The lifetime of the DM is $1.1\times 10^{25}$ sec.
The red solid line represents the signal plus background and the green dashed the background only.
The blue points represent the current PAMELA data~\cite{Adriani:2008zr}.
}
\label{fig:fraction}
\end{center}
\end{figure}

These decay products emit high energy positrons, and these positron are detected as cosmic rays.
Especially, the decays into the gaugino are rich positron source, since 
almost all of the gauginos decay into the next-to-lightest SUSY particle (NLSP) slepton emitting lepton(s),
and subsequently the slepton decays into a gravitino and a lepton.
To evaluate the positron spectrum from the decay of the baryon,
we have used the program PYTHIA~\cite{Sjostrand:2006za}.
We estimate the positron fraction following Refs.~\cite{Hisano:2005ec,Ibarra:2008qg,Chen:2008yi}. 
In Fig.~\ref{fig:fraction}, we show the positron fraction in our model.
Here we take  $m_B=100$ TeV, $m_{M}=40$ TeV and the lifetime of the DM is $1.1\times 10^{25}$ sec.
Interestingly, the positron flux from the baryonic DM decay can naturally 
explain the positron excess for $\gsim 10$ GeV  recently
reported by the PAMELA collaboration~\cite{Adriani:2008zr}.

\section{Discussion and conclusions}
A baryonic bound state of
strongly interacting sector in low scale gauge mediation,
which have mass of ${\cal O}(100)$ TeV and annihilation cross section close to the unitarity bound,
can naturally be the cold dark matter.
Such a scenario predicts an ultralight gravitino with a mass of ${\cal O}(1)$ eV,
which is attractive since it circumvents all the
notorious cosmological gravitino problems.

If the baryonic DM is composed of messenger quarks charged under strongly interacting ${\rm SU}(N)$,
the neutrality of the DM requires $N=5$.
In this letter, we pointed out that such a baryonic DM is in general metastable
due to the higher dimensional, baryon-number violating operators, and showed that
its lifetime becomes ${\cal O}(10^{24-26})$ sec.
Interestingly enough, the products of the baryonic DM decay can be observed, 
opening the possibility of its indirect detection.
We have shown that the positron flux from the baryonic DM decay can naturally 
explain the anomalous positron flux 
recently observed by the PAMELA collaboration~\cite{Adriani:2008zr}.
We should emphasize that, as shown in Fig.~\ref{fig:fraction},
the heavy baryonic DM decay predicts that the positron excess continues up to higher energy,
which can be tested by the near future PAMELA data up to about 300 GeV.

In the MSSM LSP DM scenario, 
the positron flux anomaly stretching up to higher energy indicates the heavier LSP,
which may imply the difficulty of the SUSY discovery at the LHC.
However, our present model is free from this difficulty.

In addition to the positron signal, this model also predicts high energy cosmic rays such as gamma rays and neutrinos,
which are tested by the current of future experiments. A more detailed study along this line will be given elsewhere.

\paragraph{\it Note Added:} Recently the ATIC collaboration reported an excess in the total $e^- + e^+$ flux~\cite{:2008zz}, in consistent with the PPB-BETS experiment~\cite{Torii:2008xu}.
This excess may also be explained by the composite DM decay if its mass is ${\cal O}({\rm TeV})$, 
which can be realized in our GMSB model by using the SUSY breaking effects to lower the messenger scalar masses~\cite{HNSY2}.

\section{Acknowledgement}
We would like to thank C.~Chen, T.~Moroi, H.~Murayama, M.~M.~Nojiri, F.~Takahashi and S.~Torii for useful discussion.
This work was supported by World Premier International Center Initiative (WPI Program), MEXT, Japan. 
The work of SS is supported in part by JSPS Research Fellowships for Young Scientists.

\end{document}